\documentclass[superscriptaddress, prb, reprint]{revtex4-1}

\usepackage{graphicx}
\usepackage{dcolumn}
\usepackage{bm}

\begin{document}


\title{Effective anisotropy gradient in pressure graded [Co/Pd] multilayers} 



\author{B. J. Kirby}
\address{Center for Neutron Research, NIST, Gaithersburg, MD 20899, USA}

\author{P. K. Greene}
\address{Physics Department, University of California, Davis, California 95616, USA}

\author{B. B. Maranville}
\address{Center for Neutron Research, NIST, Gaithersburg, MD 20899, USA}

\author{J. E. Davies}
\address{Advanced Technology Group, NVE Corporation, Eden Prarie, MN 55344, USA}

\author{Kai Liu}
\address{Physics Department, University of California, Davis, California 95616, USA}


\date{\today}

\begin{abstract}
A vertically graded anisotropy profile has been proposed as an optimized balance of low coercivity and thermal stability for multilayers used in magnetic media.  Deposition pressure is known to have a profound effect on the magnetic reversal properties of Co/Pd multilayers, making it an attractive control parameter for achieving an anisotropy gradient.  We have used polarized neutron reflectometry to study the depth-dependent reversal behavior of "pressure-graded" Co/Pd, and observed pronounced gradients in the saturation magnetization and in the rate at which magnetization changes with field (the effective anisotropy).  While the anisotropy gradient likely arises from a combination of factors intrinsic to deposition pressure, micromagnetic simulations indicate that the observed saturation magnetization gradient alone has a major effect on the resulting coercivity. 
\end{abstract}


\maketitle 

\section{Introduction}
The magnetic recording "trilemma" describes the frustration inherent in simultaneously optimizing signal-to-noise, maximizing thermal stability, and reducing switching field distribution in magnetic media.\cite{Richter_IEEE_1999}  In particular, the latter two goals are seemingly mutually exclusive - it is desirable to write bits with the smallest possible field, but once written, those bits need to be as stable as possible.  Exchanged coupled composites (ECC) - featuring a high anisotropy "hard" magnetic layer to serve as an anchor against thermal fluctuations exchange coupled to a low anisotropy "soft" magnetic layer to assist reversal - have been proposed as an optimized solution.\cite{Victora_IEEE_2005}  Suess took the concept further, proposing that a multilayer where the anisotropy gradually varied from top to bottom of the stack would constitute an ideally optimized ECC.\cite{Suess_APL_2006}  Since, such "graded anisotropy" magnetic multilayers have been studied both theoretically\cite{Zimanyi_JAP_2008, Suess_APL_2008, Skomski_JAP_2008, Suess_JMMM_2009} and experimentally, including for multilayers where the anisotropy profile was controlled by varying layer thickness,\cite{Kirby_PRB_2010} composition,\cite{Zha_APL_2010} and substrate temperature.\cite{Zhang_APL_2013}  
\begin{figure}
\begin{center}
\includegraphics[width = 7.cm]{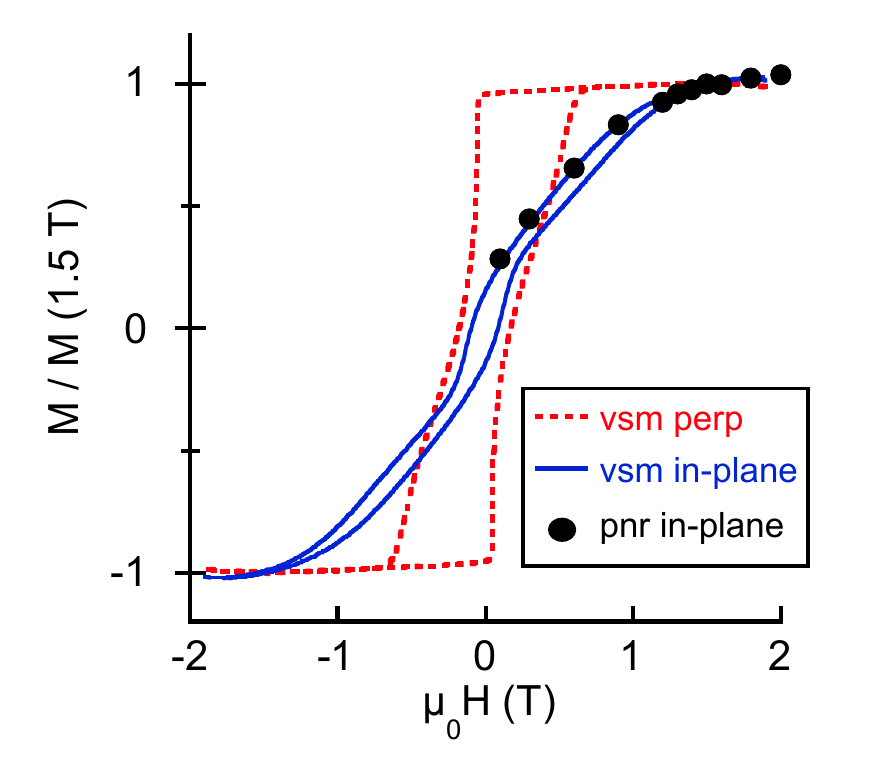}
\caption{\label{}Normalized field-dependent magnetization normalized for samples as measured with VSM, with field applied perpendicular (dashed lines) and parallel (solid lines) to the sample surface.  Solid symbols correspond to the integrated in-plane magnetization profiles as measured with PNR.}
\end{center}
\end{figure}
\begin{figure*}
\begin{center}
\includegraphics[width = 14.cm]{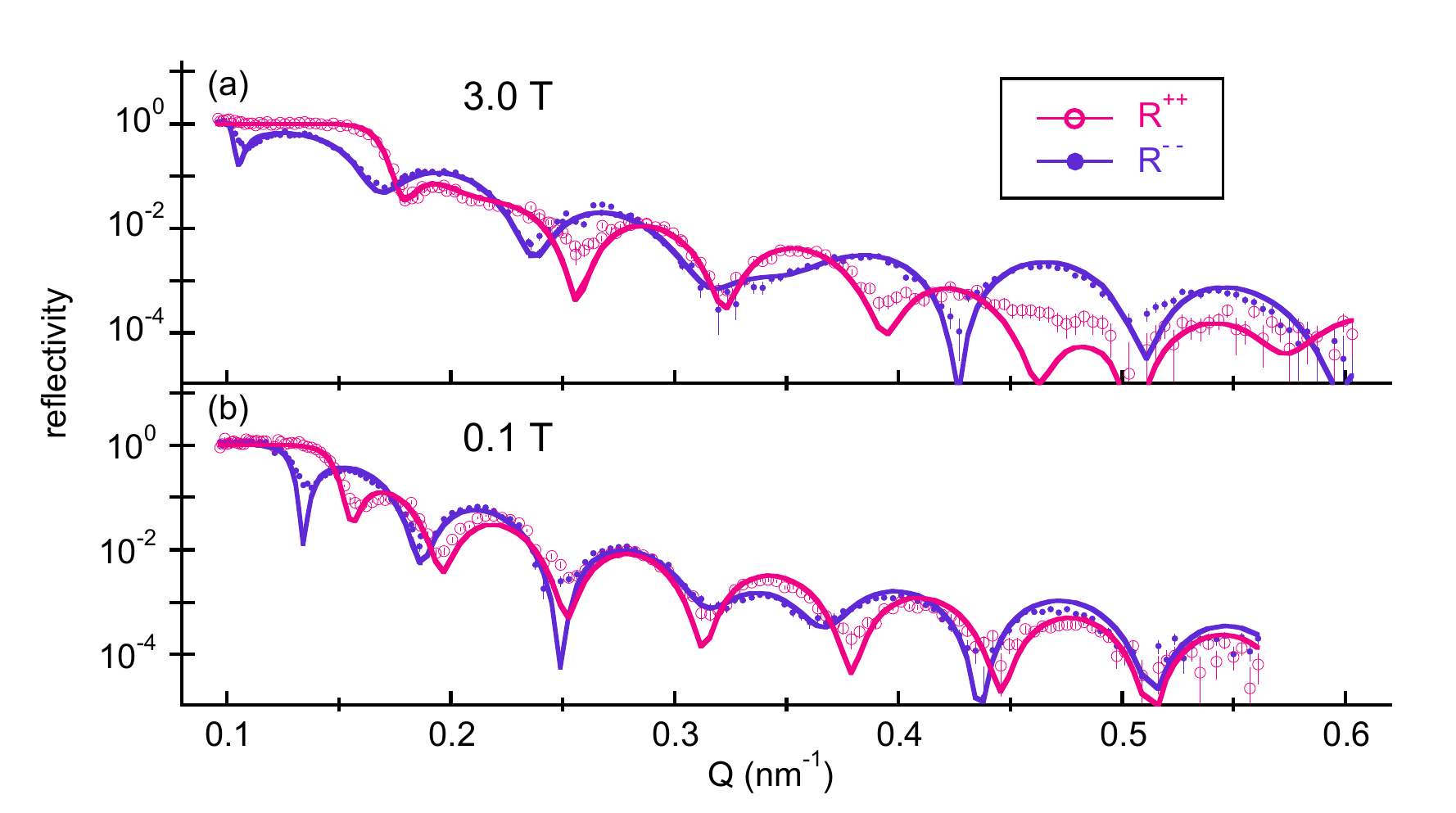}
\caption{\label{}Example reflectivities measured at (a) 3.0 T, and (b) 0.1 T.  Solid lines are model fits to the data.  Error bars correspond to $\pm$ 1 $\sigma$.}
\end{center}
\end{figure*} 

During magnetron sputtering, deposition pressure is known to have strong influence on the magnetic properties of sputtered Co/Pt \cite{Pierce_PRL_2005} and and Co/Pd\cite{Kirby_JAP_2009,Davies_JAP_2011} multilayers.  As sputtering pressure is increased, the film becomes more disordered, and reversal becomes a more localized process.  This results in smaller domains, increased coercivity, wider switching field distribution, and decreased saturation magnetization.\cite{Pierce_PRL_2005,Davies_JAP_2011} Thus, varying the sputtering pressure throughout deposition is potentially an attractively simple method for tailoring depth-dependent reversal properties without changing the material composition - for example to achieve a graded anisotropy structure.  Such "pressure-graded" Co/Pd samples have been studied with magnetometry, scanning electron microscopy with polarization analysis, and polarized neutron reflectometry (PNR),\cite{Kirby_JAP_2009,Davies_JAP_2011,Greene_APL_2014} but these studies have not addressed the nature of the anisotropy profile - i.e. the rate at which the magnetization at different depths in the multilayer changes with field.   With this in mind, we have used PNR to determine the depth-resolved hard axis magnetization curves for a pressure-graded Co/Pd multilayer.     

\section{Experiment}
Room temperature Ar$^{+}$ magnetron sputtering was used to deposit the sample onto a Si (100) substrate.  The base pressure of the chamber was 1.33 $\mu$Pa.  A 20 nm Pd seed layer was sputtered at an argon pressure of 0.7 Pa, followed by 30 bilayer repeats of [0.4 nm Co / 0.6 nm Pd] at 0.7 Pa, 15 bilayer repeats at 1.6 Pa, 15 bilayer repeats at 2.7 Pa, and a 4.4 nm Pd cap layer at 2.7 Pa. Vibrating sample magnetometry (VSM) was used to measure the room temperature field-dependent magnetization, for magnetic field applied both perpendicular and parallel to the plane of the sample, as shown in Figure 1.  The perpendicular hysteresis loop is much more square than the in-plane, indicating that \em collectively \em, the sample exhibits perpendicular magnetic anisotropy.

Room temperature PNR measurements were used to characterize the properties of the individual pressure regions.  Primary measurements were taken using Asterix at the Los Alamos Neutron Science Center, while supporting measurements were conducted using the NG-1 Reflectometer at the NIST Center for Neutron Research.  PNR is sensitive to the depth profiles of the nuclear composition and in-plane magnetization of thin films and multilayers.\cite{Fitz_book,Chuck_book}  Specifically, for neutrons with magnetic moment polarized either parallel (+) or anti-parallel (-) to an applied magnetic field $H$, the non spin-flip wavevector transfer-dependent specular reflectivities $R(Q)^{++}$ and $R(Q)^{--}$ are functions of the spin-dependent scattering length density depth ($z$) profiles, 
\begin{eqnarray}
\rho ^{++}(z) = \rho_{N} + CM,
\end{eqnarray}
\begin{eqnarray}
\rho ^{--}(z) = \rho_{N} - CM.
\end{eqnarray}
where $\rho_{N}$ is indicative of the nuclear composition, $M$ is the in-plane projection of the sample magnetization parallel to the applied field, and $C$ is a constant.\footnote{$C$ = 2.853 x 10$^{-7}$, for $M$ in kA m$^{-1}$ and $\rho$ in nm$^{-2}$.} 
It is straightforward to exactly calculate the reflectivity corresponding to a given scattering length density profile,\cite{Chuck_book} thus depth profiles of both $\rho$ and $M$ can be determined through model fitting of the measured spin-dependent reflectivities.  Since specular PNR is insensitive to the component of the magnetization perpendicular to the sample surface,\cite{Chuck_book} primary measurements were conducted as a function of decreasing in-plane $H$ from 3.0 - 0.1 T, followed by a measurement at 10 T.  This procedure allowed for characterization of the in-plane magnetization depth profile as spins in the sample progressively relaxed away from a hard-axis orientation.  Note that spin-flip scattering ($R^{+-}$ and $R^{-+}$) is not to be expected in this geometry, as it arises from the in-plane component of the sample magnetization perpendicular to $H$.  This was experimentally confirmed by a limited number of measurements with a spin analyzer between the sample and detector.  Examples of the fitted PNR data measured at 3.0 T and 0.1 T are shown in Figure 2.  

From Eq. 1-2, the sample magnetization is manifest in the differences between $R^{++}$ and $R^{--}$.  Fig. 2 shows clear spin-dependence in the reflectivities at both high and low field, indicating sensitivity to the magnetic depth profile.  Further, significant changes are observed as the field is reduced, indicating sensitivity to the reduction in in-plane magnetization.  Solid lines in Fig. 2 are fits to the data generated using the Refl1D software package.\cite{Kirby_COCIS_2011}  The fits reproduce the data well, and correspond to nuclear and magnetic depth profiles shown in Figure 3.  
\begin{figure}
\begin{center}
\includegraphics[width = 7cm]{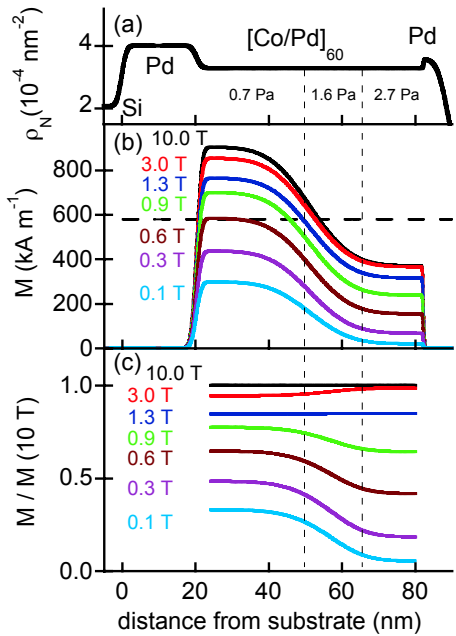}
\caption{\label{}  Depth profiles determined from model fitting of the PNR data.  (a) Nuclear scattering length density depth profile.  (b) Depth profiles of the in-plane magnetization component parallel to the applied field shown at several example fields.  The horizontal dashed line corresponds to the maximum theoretical magnetization of [0.4 nm Co / 0.6 nm Pd] if the magnetization arises solely from Co.  (c) Depth profiles shown in (b) normalized by the 10 T saturation profile.}
\end{center}
\end{figure}
The $z$-dependent nuclear scattering length density used to model the data at all fields is shown in Fig. 3 (a), with features corresponding to the Si substrate, Pd seed layer, Co/Pd multilayer and Pd cap. As the $Q$-range measured was insufficient to resolve the individual Co and Pd layers, and since neither the density or relative composition is expected change significantly with deposition pressure, the Co/Pd multilayer is modeled as CoPd "alloy" with constant $\rho_{N}$.\cite{Kirby_PRB_2010}(note, allowing $\rho_{N}$ to vary provided no significant improvement in fit quality).  Vertical dashed lines in Fig. 3 indicate the expected positions of the three pressure regions.  

The in-plane magnetization depth profiles at selected fields are shown in Fig. 3(b).  These profiles can be compared to VSM results by integrating $M$ over all $z$, as shown by the solid points in Fig. 1.  The normalized integrated values agree well with the corresponding VSM measurements, a strong confirmation of the model fitting. In contrast to the flat Co/Pd nuclear profile, a highly non-uniform magnetic profile was required to fit the data.  The magnetic model used consists of two magnetic layers with a broad ~ 30 nm error function interface.  This model was found to fit the data better than 3 discrete layers with sharp interfaces, even at the highest applied fields.  The 2 $\sigma$ uncertainty for the high and low $z$ magnetizations was calculated using a Markov chain Monte Carlo algorithm,\cite{ISO, dream, Kirby_COCIS_2011} and was found to be less than 10 kA m$^{-1}$ for all values. 

At the initial applied field of 3 T, the magnetization profile is highly nonuniform, as $M$ decreases with increasing $z$.  The similar profile determined from a subsequent measurement at 10 T shows that this nonuniformity is not due to lack of magnetic saturation, but instead originates from a saturation magnetization ($M_{S}$) that varies significantly with $z$.  For high field, the magnetization of the low pressure region is too large to be attributed to Co alone, implying a contribution due to Pd polarization near Co interfaces.\cite{Carcia_SM_1985}  Assuming a Co $M_{S}$ = 1444 kA m$^{-1}$,\cite{Fitz_book} and  40\% Co concentration, the maximum $M$ arising purely from Co should be 578 kA m$^{-1}$, denoted by the horizontal dashed line in Fig. 3(b).  Attributing the excess magnetization to interfacial polarization implies a Pd areal magnetization of up to 0.3 mA (3 $\times10^{-5}$ emu cm$^{-2}$) in the low pressure region, comparable to values determined previously for epitaxial Co/Pd superlattices.\cite{Engel_JAP_1991,Engel_PRL_1991}

As field is reduced, $M$ drops for all $z$, corresponding to spins rotating away from the hard axis direction.  While the magnetization gradient is apparent at all fields in Fig. 3(b), the depth-dependent variation in $M_{S}$ makes it difficult to discern variations in anisotropy - i.e. it masks variations in the \em rate \em at which the $M$ drops with $H$.  To make this rate apparent, Figure 3(c) shows the magnetization profiles in 3(b) normalized by the nominally saturating 10 T profile.  Plotting in this way, we see that particularly at low $H$, $M$ drops with reduced $H$ much faster at high $z$ than at low $z$.  i.e., the sample is progressively harder to magnetize in-plane as deposition pressure increases with increasing distance from the substrate interface.  These field-dependent profiles can be put into a more familiar context by extracting slices of the field-dependent magnetization profile, and plotting $M(H)$ corresponding to different depths in the sample.  Figure 4(a) shows $M(H)$ for regions deposited at 0.7 Pa ($z$ = 24 nm), 1.6 Pa ($z$ = 56 nm) and 2.7 Pa ($z$ = 80 nm), while Figure 4(b) shows the ratio of the 0.7 Pa region magnetization to that of the 2.7 Pa and 1.6 Pa regions.\begin{figure}
\begin{center}
\includegraphics[width = 7cm]{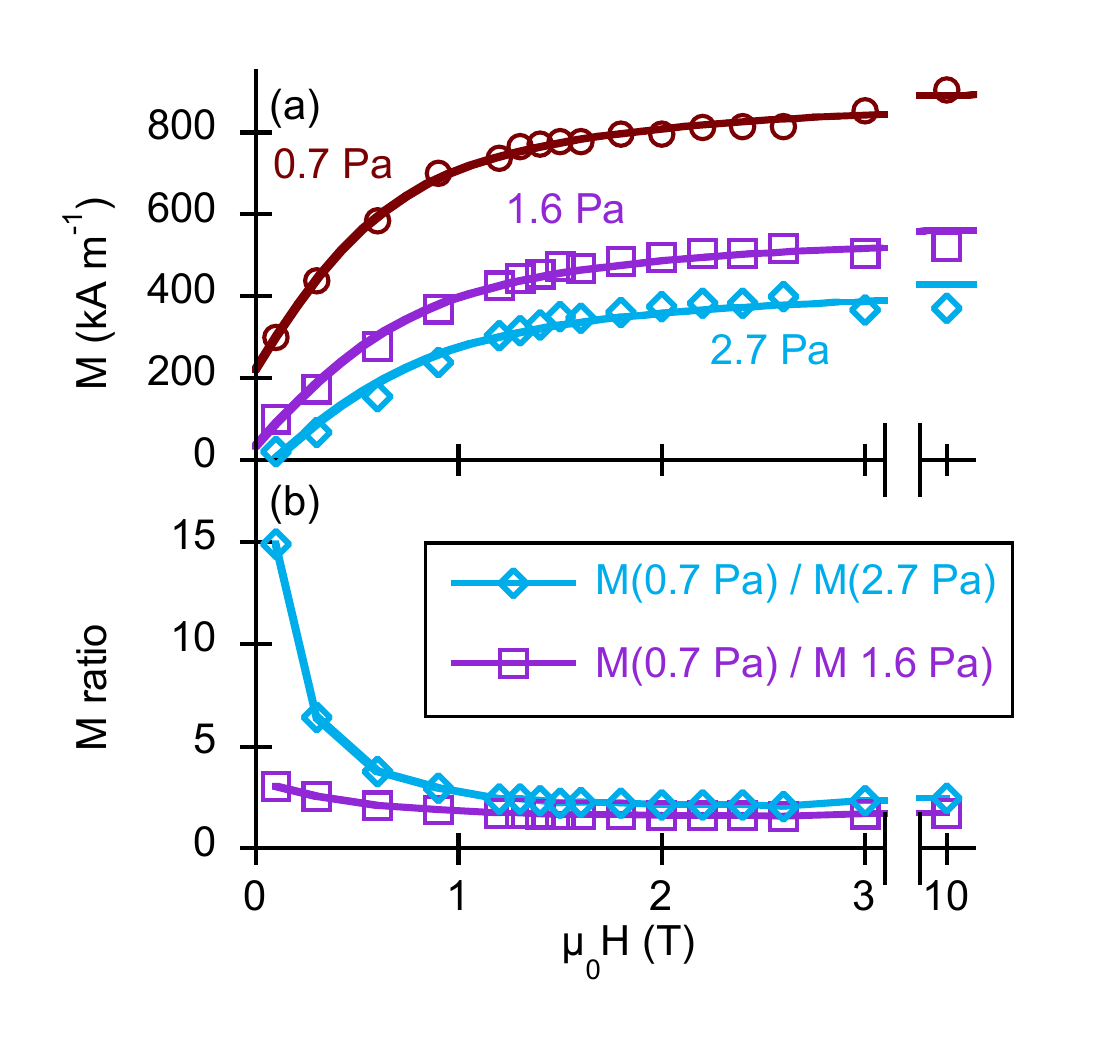}
\caption{\label{}(a) Field-dependent magnetizations of regions deposited at different pressures (depths). (b) Ratio of the 0.7 Pa region magnetization to the magnetization in the 2.7 Pa and 1.6 Pa regions, respectively.  Solid lines are guides to the eye. }
\end{center}
\end{figure}
Fig. 4(a) demonstrates the clear pressure (depth) dependent variations in saturation magnetization, while Fig. 4(b) shows that as as pressure is increased the magnetization drops faster with decreasing field.   That the rate at which magnetization drops with field indeed changes with depth, demonstrates that pressure grading does result in a corresponding gradient in the \emph{effective} anisotropy.  

\section{Discussion}
The observed gradient in effective anisotropy likely arises from several factors that are difficult to disentangle, including pressure-dependent variations in crystalline disorder, intrinsic magnetocrystalline and interfacial anisotropy, domain size, and saturation magnetization.  While we cannot uniquely isolate the depth-dependencies of the rest of these parameters, PNR measurements do reveal the depth-dependence of $M_{S}$.  This $M_{S}$ gradient plays a critical role in the reversal behavior, and must be accounted for in potential device applications.  Consider that to first order,  the \emph{field} associated with uniaxial anisotropy of energy density (anisotropy constant) $K$ is linearly dependent on both $K$ and $M_{S}$\cite{Chikazumi} 
\begin{eqnarray}
H_{A} = \frac{2K}{\mu_{0}M_{S}}.
\end{eqnarray}
Therefore, by definition, the observed 40 \% gradient in $M_{S}$ linearly affects the profile of the anisotropy field.  In addition, $M_{S}$ (i.e. the magnitude of the magnetization vector) has a linear effect on the Zeeman energy - the energy penalty corresponding to the deviation of magnetization vector from the direction of the applied field. If we define $\phi$ to be the angle between the easy axis and the magnetization vector, the Zeeman energy is defined\cite{Chikazumi} 
\begin{eqnarray}
W_{z} = \mu_{0} H M_{S}\cos{(\phi-\frac{\pi}{2}}).
\end{eqnarray}
\begin{figure}
\begin{center}
\includegraphics[width = 7cm]{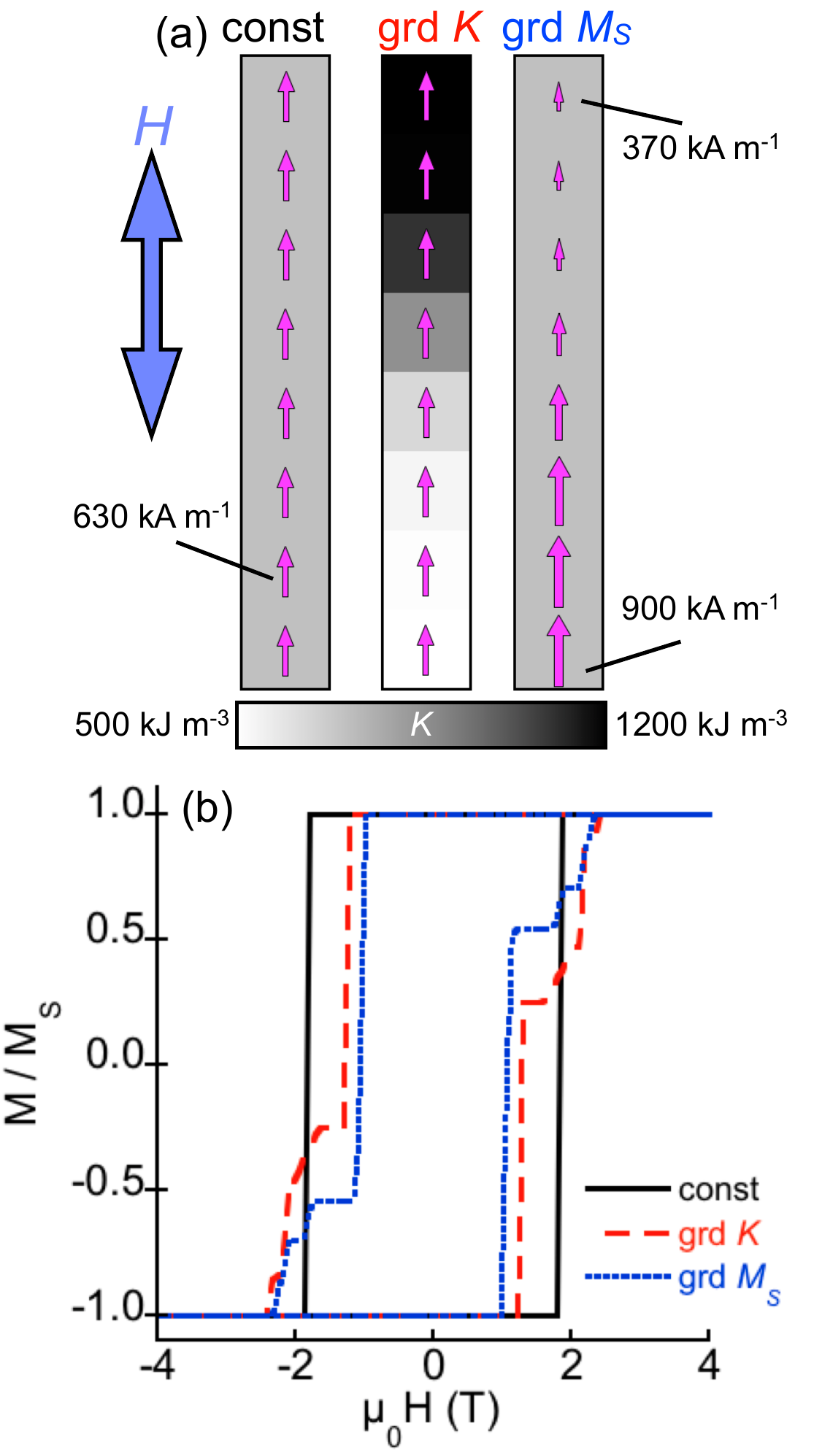}
\caption{\label{}(a) Cartoon depiction of spin structures used for micromagnetic simulations.  All have the same average $K$ and $M_{S}$, but different $K$ and $M_{S}$ depth profiles.   From left: constant $K$ with constant $M_{S}$, graded $K$ with constant $M_{S}$, and constant $K$ with graded $M_{S}$.  Magnitude of $M_{S}$ is depicted by arrow size, while magnitude of $K$ is depicted by grayscale.  (b) Simulated hysteresis loops for $H$ along the long axis for the three pillars shown in (a). }
\end{center}
\end{figure}
Thus, we identify two simple channels for $M_{S}$ to linearly affect reversal behavior.  To further illustrate this point, we have used the OOMMF micromagnetic software package\cite{oommf} to simulate easy-axis hysteresis loops for three different "pillars" of spins, all with the same \emph{average} anisotropy constant and saturation magnetization, but with different \emph{distributions} of $K$ and $M_{S}$.  Each pillar consists of a 10 $\times$ 10 $\times$ 64 nm array of 1 nm$^{3}$ spins, with an exchange constant of $A$ = 0.178 pJ m$^{-1}$, a value chosen such that the spins in different regions are neither rigidly coupled, nor effectively decoupled.  The average $M_{S}$ = 629 kA m$^{-1}$ (i.e. the average value determined from PNR), and the average $K$ = 710 kJ $m^{-3}$ (a reasonable value for Co/Pd\cite{Manchanda_JAP_2008}).     Cartoon depictions of the three spin structures considered are shown in Figure 5(a):  constant $K$ (710 kJ m$^{-3}$) with constant $M_{S}$ (629 kA m$^{-1}$), a 40 \% graded $K$ (500-1200 kJ m$^{-3}$) with constant $M_{S}$ (629 kA m$^{-1}$), and constant $K$ (710 kJ m$^{-3}$) with a 40 \% graded $M_{S}$ (370-900 kA m$^{-1}$, i.e. the $M_{S}$ profile shown for Fig. 3(c)).  Figure 5(b) shows simulated hysteresis loops for $H$ along the long (easy) axes of the pillars.  In this example, a 40 \% gradient in $K$ indeed softens the loop, resulting in a 19 \% reduction in coercive field as compared to the constant $K$, constant $M_{S}$ pillar.  However,  a 40 \% gradient in $M_{S}$ results in an even larger 29 \% decrease in coercive field.  While the actual mechanism of reversal in pressure-graded Co/Pd is certainly more complicated than can be described with this simple simulation, this result emphasizes the significant role of the $M_{S}$ gradient.  This intrinsic variation in $M_{S}$ certainly must be considered in the design of actual devices based on pressure-graded Co/Pd, and may be exploitable as a knob for precisely tailoring magnetization reversal properties in magnetic media.

\section{Conclusion}
We explicitly demonstrate that a simple technique of varying pressure during magnetron sputtering can be used to create magnetic multilayers exhibiting a pronounced vertical gradient in the effective anisotropy.  In addition to the anisotropy gradient, we find that pressure grading leads to a pronounced gradient in the saturation magnetization, and that this saturation magnetization gradient critically affects the magnetic reversal properties.  Further, this work demonstrates how detailed polarized neutron reflectometry measurements of the depth-resolved, hard-axis magnetization curves can be used to extract information about the depth-dependent anisotropy that is difficult or impossible to acquire through other techniques.    Support from the NSF Materials World Network program (DMR-1008791) is gratefully acknowledged. We are extremely grateful to M. R. Fitzsimmons of Los Alamos National Laboratory for assistance with Asterix, and thank Randy K. Dumas of Gothenburg University, and J. A. Borchers and P. A. Kienzle, of NIST for valuable discussions. 

\providecommand{\noopsort}[1]{}\providecommand{\singleletter}[1]{#1}%

\end{document}